\documentclass[aps,superscriptaddress,twocolumn,showpacs,secnumarabic,a4paper,floatfix]{revtex4}

\usepackage{amsmath}
\usepackage{amssymb}
\usepackage{color}
\usepackage{graphicx}% Includefigure files
\usepackage{dcolumn}% Align table columns on decimal point
\usepackage{bm}
\usepackage{amsbsy}

\newcommand{\nvec}{\mathbf{n}}

\newcommand{\lvec}{\mathbf{l}}
\newcommand{\mvec}{\mathbf{m}}

\newcommand{\Qvec}{\mathbf{Q}}
\newcommand{\Qij}{Q_{ij}}

\DeclareMathOperator{\Tr}{Tr}

\begin{document}

\author{Z. Eskandari}
\affiliation{Centro de F{\'\i}sica Te\'orica e Computacional,
Universidade de Lisboa, Avenida Professor Gama Pinto 2, P-1649-003 Lisboa, Portugal}

\author{N. M. Silvestre}
\email[]{nunos@cii.fc.ul.pt}
\affiliation{Departamento de F{\'\i}sica da Faculdade de Ci\^encias,}
\affiliation{Centro de F{\'\i}sica Te\'orica e Computacional,
Universidade de Lisboa, Avenida Professor Gama Pinto 2, P-1649-003 Lisboa, Portugal}

\author{M. Tasinkevych}
\affiliation{Max-Plank-Institut f\"ur Intelligente Systeme, Heisenbergstr. 3, D-70569 Stuttgart, Germany.}
\affiliation{Institute f\"ur Theoretische und Angewandte Physik, Universit\"at Stuttgart, Pfaffenwaldring 57, D-70569 Stuttgart, Germany.}
\email[]{miko@is.mpg.de}

\author{M. M. Telo da Gama}
\affiliation{Departamento de F{\'\i}sica da Faculdade de Ci\^encias,}
\affiliation{Centro de F{\'\i}sica Te\'orica e Computacional,
Universidade de Lisboa, Avenida Professor Gama Pinto 2, P-1649-003 Lisboa, Portugal}

\title{Interactions of distinct quadrupolar nematic colloids}

\begin{abstract}

The effective interaction between spherical colloids in nematic liquid crystals is investigated in the framework of the Landau-de Gennes theory. The colloids differ through their interaction with the nematic. While both particles induce quadrupolar far-field distortions in the nematic matrix, with unlike quadrupole moments, one favours homeotropic and 
the other degenerate planar anchoring of the nematic director. 
In the strong anchoring regime the colloids with homeotropic anchoring are accompanied by an equatorial disclination line defect, known as ``Saturn-ring'', while the colloids with 
degenerate planar anchoring nucleate a pair of antipodal surface defects, called ``Boojums''. In the linear (large-distance) regime the colloidal interactions are of the quadrupolar type, where the quadrupoles have opposite signs. These are attractive when the colloids are aligned either parallel or perpendicular to the far-field director.
At short distance, non-linear effects including ``direct'' interactions between defects give rise to a repulsion between the particles, which prevents them from touching.
This finding supports the stability of nematic colloidal square crystallites the assembly of which has been reported recently.
\end{abstract}

\pacs{61.30.-v, 61.30.Jf, 82.70.Dd}
\keywords{ nematic liquid crystals, colloidal interactions}%Use showkeys class option if keyword

\maketitle

\section{Introduction}

The self-assembly of colloidal particles into structures with controlled spatial ordering is of great importance in colloid science, with particular interest in the assembly of photonic crystals \cite{Musevic.2011} -- artificially produced periodic dielectric structures designed to control and manipulate light. In this context, a variety of colloidal structures assembled in liquid crystal (LC) matrices \cite{Musevic.2006,Ravnik.2007}, combined with the unique mechanical and electro-optical properties of the LC host \cite{Kang.2001}, have proven to be good candidates for the development of colloid crystals with tunable photonic properties.

In conventional colloids, in isotropic fluids, the colloidal particles interact via van der Waals, electrostatic, or steric forces. These forces are isotropic, and their range does not exceed a few tens of nanometers. By contrast, when dispersed in a LC, due to its long-range orientational molecular ordering, colloidal particles interact predominantly through long-range anisotropic forces \cite{Stark.2001,Tasinkevych.2010}. The origin of these effective forces is the elastic distortions of the LC matrix due to the presence of the colloidal particles. The range of the elastic forces is of the order of several colloidal diameters. 
Elastic forces drive the particles to self-assemble into linear chains \cite{Loudet.2000,Cluzeau.2001,Cluzeau.2002a,Voltz.2004}, periodic lattices \cite{Voltz.2004, Nazarenko.2001,Cluzeau.2002}, anisotropic clusters \cite{Poulin.1999}, and cellular structures \cite{Meeker.2000}.

A distinctive feature of LC colloids is the presence of topological defects \cite{Mermin.1979}, that not only determine the symmetry of the long-range colloidal interaction \cite{Lubensky.1998,Lev.1999}, but also stabilize the ordered aggregates at short range \cite{Musevic.2006}, where the elastic interactions are dominated by non-linear effects and render 
the self-assembly of LC colloids a challenging theoretical problem. Topological defects in nematic liquid crystals (NLC) are nucleated due to the mismatch of the global and local (at the colloidal surfaces) boundary conditions leading to frustration of the uniform nematic order. 

Small spherical particles ($\lesssim1\mu$m) imposing homeotropic surface anchoring on the nematic director stabilize equatorial Saturn-ring defects \cite{Gu.2000}, which for larger particles may be stabilized through confinement \cite{Skarabot.2008}, or by external electric fields \cite{Loudet.2001}. The far-field distortions have quadrupolar symmetry and the resulting pairwise colloidal interaction is of the quadrupolar type decaying with the distance $d$ between the particles as $d^{-5}$ \cite{Lubensky.1998}. 

Larger particles ($\gtrsim {\cal O} (1\mu$m)) with homeotropic anchoring induce point-like hedgehog defects \cite{Poulin.1998}, which lead to a far-field director of dipolar symmetry and a large distance colloidal interaction varying as $d^{-3}$ \cite{Lubensky.1998}. For particles with planar degenerate anchoring, two antipodal surface defects, known as boojums \cite{Poulin.1998} are nucleated which lead to a far-field director of quadrupolar symmetry. Recent numerical calculations reveal that the cores of nematic boojums can take three different 
configurations \cite{Silvestre.inprep}: single core, double core, or split core. The single-core boojum is a point-like index $1$ defect with azimuthal symmetry, the split-core boojum has two index $1/2$ surface point-like defects connected by a $1/2$ bulk disclination line. The double-core boojum is an intermediated structure with broken azimuthal symmetry. The far-field distortions and the resulting asymptotic pair interaction between particles with degenerate planar anchoring are of the quadrupolar type \cite{Pergamenshchik.2010}, as for
Saturn-ring particles.

Recently, a light sensitive coating of colloidal particles was used in order to switch the surface anchoring from homeotropic to planar, reversibly \cite{Chandran.2011}, providing the means for tuning the colloidal interactions and thus controlling the assembly of the colloidal structures.

Direct assembly of two-dimensional $(2D)$ crystallites of quadrupolar \cite{Skarabot.2008}, dipolar \cite{Skarabot.2007}, or dipolar and quadrupolar \cite{Ognysta.2009}, spherical particles has been achieved by using laser tweezer techniques. The crystallites are stabilized by the presence of topological defects, which provide local free energy minima of the order of $\sim1000k_BT$, where $k_B$ is the Boltzmann constant and $T$ the absolute temperature. By contrast, interacting quadrupolar boojum-particles in three-dimensional systems do not exhibit short-range repulsive behaviour, and the equilibrium configuration corresponds to close contact or coalescence of the particles \cite{Smalyukh.2005a,Mozaffari.2011,Silvestre.inprep}. 
\begin{figure}[t]  %{r}{8cm} % "l" or "r" for the side on the page.
\centerline{\includegraphics[width=.5\columnwidth]{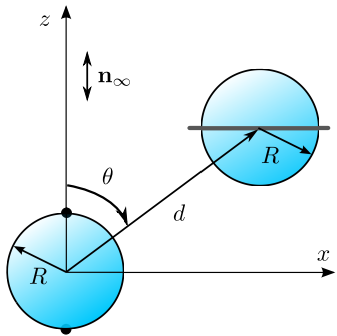}}
\caption
{Schematic representation of interacting Saturn-ring and boojum quadrupolar particles of radius $R$. $d$ is the inter-particle distance and $\theta$ is the polar angle relative to the far-field director.}
\label{dx}
\end{figure}

Recently, the assembly of $2D$ colloidal crystallites of spherical particles with two types of anchoring, homeotropic and degenerate planar, dispersed in 5CB has been reported \cite{Ognysta.2011}. The particles with radius $R=2.16\mu$m were confined to a cell of thickness $6\mu$m in order to stabilize the Saturn-ring configuration around the particles with homeotropic anchoring. Both types of particles generate quadrupolar nematic distortions, but the corresponding quadrupolar moments have different signs, meaning that the particles attract each other when they are aligned parallel or perpendicular to the far-field director. This allowed the assembly of $2D$ square colloidal crystallites. In the following we shall adopt the notations of Ref.~\cite{Ognysta.2011}: a Saturn-ring quadrupolar particle will be denoted by ``S'', and a boojum quadrupolar particle by ``P'' (planar anchoring). In this article, we present the results of a numerical study of the NLC-mediated interaction between S and P particles (see Fig.~\ref{dx}) for a wide range of distances. We focus on the short-distance regime that determines the final equilibrium configuration and its stability.

The paper is organized as follows: in the next section we introduce the Landau-de Gennes free energy functional, which we minimize numerically. In Sec.~\ref{sec3} we present our results. We show that the S-P interaction exhibits two local minima for configurations in which particles align either parallel or perpendicular to the far-field director, thus allowing the assembly of square lattices. We show that the two equilibrium configurations are separated by a free energy barrier of the order of
$250 k_B T (R/1 \mu \mbox{m})$, where $R$ is the colloidal radius. In Sec.~\ref{conclusions} we present our conclusions.

\section{Landau-de Gennes Theory}
\label{sec2}

Phenomenologically, nematic liquid crystals are characterised by a traceless, symmetric tensor order-parameter $\Qvec$ which can, in general, be written as \cite{Vissenberg.1997}

\begin{equation}
\Qij=\frac{Q}{2}\left(3n_i n_j-\delta_{ij}\right)+\frac{B}{2}\left(l_il_j-m_im_j\right),
\end{equation}
where $Q$ is the uniaxial order parameter, which measures the degree of orientational order along the nematic director $\nvec$, and $B$ is the biaxial order parameter, which measures the degree of orientational order along the directions perpendicular to $\nvec$, characterised by $\lvec$ and $\mvec$.
The corresponding Landau-de Gennes (LdG) free energy functional is \cite{Gennes.1993}

\begin{equation}
F=\int_\Omega{d^3x\left(f_b+f_e\right)}+\int_{\partial\Omega}{ds\,f_s},
\label{LdG}
\end{equation}
where $f_b$ and $f_e$ are the bulk and elastic free energy densities, given by

\begin{eqnarray}
f_b&=&a(T)\Tr\Qvec^2 -b\Tr\Qvec^3 +c\left(\Tr\Qvec^2)\right)^2, \label{f_bulk}\\
f_e&=&\frac{L}{2}\left(\partial_k\Qij\right)^2. \label{f_elastic}
\end{eqnarray}
The first integral in Eq.(\ref{LdG}) is over the volume occupied by the nematic, $\Omega$, and the second is over the surfaces of the colloidal particles, $\partial\Omega$. The bulk parameter $a(T)=a_0\left(T-T^*\right)$ depends linearly on the temperature $T$, with $a_0$ a material dependent constant and $T^*$ the supercooling temperature of the isotropic phase; $b$ and $c$ are positive material dependent constants. For a given temperature $T$ and in the absence of elastic distortions the free energy density in Eq.~(\ref{f_bulk}) is minimized for the (bulk) degree of orientational order given by
\begin{equation}
Q_b=\frac{b}{8c}\left(1+\sqrt{1-\frac{8\tau}{9}}\right),
\end{equation}
where $\tau=24a_0(T-T^*)c/b^2$ is the reduced temperature.
$\tau$ controls the stability of the nematic and isotropic phases. For $\tau>9/8$ the nematic phase is unstable, while the isotropic phase is unstable for $\tau<0$. The nematic and the isotropic phases coexist at $\tau=1$. For simplicity, we consider the one-elastic constant approximation where $L$ is related to the Frank-Oseen (FO) elastic constant $K$ by $K=9Q^2L/2$ \cite{Vissenberg.1997}. 

We consider the interaction between colloidal particles with homeotropic and degenerate planar anchoring. We assume rigid homeotropic boundary conditions, while the planar degenerate anchoring is described by the covariant surface anchoring free energy proposed by Fournier and Galatola (FG):

\begin{equation}
f_s=W_1\left(\tilde{Q}_{ij}-\tilde{Q}^\perp_{ij}\right)^2
+W_2\left(\tilde{Q}^2_{ij}-(3Q_b/2)^2\right)^2,
\end{equation}
$\tilde{Q}_{ij}=\Qij+Q_b\frac{\delta_{ij}}{2}, \tilde{Q}^\perp_{ij}=\left(\delta_{ik}-\nu_i\nu_k\right)\tilde{Q}_{kl}\left(\delta_{lj}-\nu_l\nu_j\right)$, and 
$\mbox{\boldmath{$\nu$}}$ is the surface normal. The quadratic term favours tangential orientation of the director $\nvec$, and the quartic term guarantees the existence of a minimum for the scalar order parameter at the surface equal to its bulk value $Q_b$. For simplicity we will assume $W_1=W_2=W$. The equilibrium surface orientation is determined by the free energy functional 
minimum that  satisfies the far-field $\nvec_\infty$ boundary conditions.

Typical values of bulk parameters for 5CB are $a_0=0.044\times10^6$J/Km$^3$, $b=0.816\times10^6$J/m$^3$, $c=0.45\times10^6$J/m$^3$, $L=6\times10^{-12}$J/m, and $T^*=307$K. At the nematic-isotropic
coexistence the bulk correlation length $\xi=\left(24 c L /b^2\right)^{1/2}\simeq 10$nm determines
the order of magnitude of the spatial extension of the cores of the topological defects \cite{Chandresakhar.1992}.
Calculations were performed for particles of radius $R=0.1\mu$m$\simeq 10\xi$, in a cubic box of length $l=30 R$;
the reduced temperature  $\tau \simeq 0.16$, and the anchoring strength $W$ satisfying
$W Q_b^2 R/K \simeq 37$, which corresponds to the strong anchoring regime.

The Landau-de Gennes free energy Eq.(\ref{LdG}) is minimized by using finite elements methods, with adaptive meshing. The surfaces of the spherical particles $\partial \Omega$ are triangulated using the open source \textit{GNU Triangulated Surface Library} \cite{GTS}. Then, the triangulation of the nematic domain $\Omega$
is carried out using the \textit{Quality Tetrahedral Mesh Generator} \cite{tetgen}, which supports
the adaptive mesh refinement. Linear triangular and tetrahedral elements are used in $2D$ and $3D$,
respectively. Generalized Gaussian  quadrature rules for multiple integrals \cite{cub_encycl} are
used in order to evaluate integrals over the elements. In particular, for tetrahedra
a fully symmetric cubature rule with 11 points \cite{keast1986} is used,  
and integrations over triangles are done by using a fully symmetric quadrature rule with 7 points \cite{stroud1971}. The discretized Landau-de Gennes functional in then minimized using the \textit{INRIA}'s \textit{M1QN3} \cite{M1QN3} optimization routine, which implements a limited memory quasi-Newton 
technique of Nocedal \cite{nocedal1980}.

\section{Results}
\label{sec3}

Lets us assume that the far-field director $\nvec_\infty$ is parallel to the $z-$axis, and that
S, or P particle is placed at the origin of the reference frame. For spherical particles,
the distortions of the director field ${\bf n}({\bf r})$ are uniaxial, with the symmetry group
$C_{\infty v}$. At large distances $r$ from the particle, the director exhibits small distortions from its uniform far-field alignment ${\bf n}({\bf r})\simeq
(n_1,n_2,1-{\cal O} (n_1^2,n_2^2))$. 
In the one-elastic constant approximation, the transverse components
$n_i\,(i = 1,2)$ satisfy Laplace's equation $\bigtriangleup n_i = 0$ \cite{Pergamenshchik.2010}, 
and the asymptotic solution for $n_i$ can be expanded in terms of multipoles. Due to the symmetry
requirements, the quadrupolar term is the lowest-order term in the expansions \cite{Pergamenshchik.2010},
 i.e.,
\begin{equation}
 n_i = 5 \sum_{\alpha,\beta=1}^{3} Q_{i\alpha\beta}\frac{r_{\alpha}r_{\beta}}{r^5}+...,
\label{multipole}
\end{equation}
and the symmetry of the director requires the quadrupole moment tensor 
$Q_{i\alpha\beta} = Q\delta_{i\alpha}\delta_{3\beta}$. In Ref.~\cite{Pergamenshchik.2010}
an explicit expression for  $Q_{i\alpha\beta}$ in terms of the surface integrals 
involving $n_i$ has been given. 
Applying then a superposition approximation, the pair interaction potential between quadrupolar particles is obtained \cite{Pergamenshchik.2007,Pergamenshchik.2010}
\begin{equation}
F_{1Q-1Q}=\frac{80\pi K}{9d^5}Q_1Q_2\left(9-90\cos^2\theta+105\cos^4\theta\right),
\label{QQ}
\end{equation}
where the subscript indicates that the interaction is between two uniaxial quadrupoles
with quadrupole moments $Q_1, Q_2$. 
The interaction potential in Eq.~(\ref{QQ}) is repulsive at $\theta = 0$ and $\theta = \pi/2$,
and attractive at intermediate orientations. The effective interactions between two S,
or two P particles have been analysed theoretically  beyond the superposition approximation in
Refs.~\cite{Tasinkevych.2010,Mozaffari.2011,Silvestre.inprep}. In both cases,
significant deviations from the asymptotic quadrupolar behaviour have been found at 
short distances.

The interaction between S and P particles was measured recently \cite{Ognysta.2011}.
Asymptotic analysis suggests that S particles induce elastic quadrupoles with quadrupole moment $Q_S>0$. 
By contrast, P particles are characterised by quadrupole moments $Q_P<0$ \cite{Pergamenshchik.2010}. The sign of a quadrupole moment cannot be obtained from the interaction of 
like quadrupoles, S-S or P-P, and only its absolute strength can be extracted from such experiments. Different experiments yield a range of values, as presented in Table~\ref{table1}. 
The discrepancies are most likely due to the different conditions (e. g. geometrical confinement) under which the experiments were performed \cite{Vilfan.2008}, or to the effects of 
thermal fluctuations that are important at large distances \cite{Smalyukh.2005a}. These factors may limit (severely) the range of validity of the asymptotic expression, Eq.\eqref{QQ}.

\begin{table}
\begin{center}  
\caption{The absolute values of the elastic quadrupole moment $|Q_P|$  as reported in different experiments for the case
of spherical colloidal particles with planar degenerate anchoring.}
\label{table1}
\begin{tabular}{  m{0.38\linewidth}   m{0.22\linewidth}   m{0.22\linewidth}   m{0.1\linewidth} }
    \hline  \hline 
     cell thickness ($\mu$m) & $2R$ ($\mu$m) & $|Q_P|$ ($R^3$) &  Ref. \\
     \hline
      8 & 4.5 & 0.17 & \cite{Kotar.2006} \\
   %  \hline
     6.5--8 & 4 & $0.2$ & \cite{Vilfan.2008} \\
   %  \hline
     6 & $4.32$ & $0.5$ & \cite{Ognysta.2011}  \\ 
    %  \hline
     30--100 &  3--$7.5$ & $0.66$ & \cite{Smalyukh.2005a} \\
    \hline  \hline 
  \end{tabular}
\end{center}
\end{table}

Recent numerical calculations revealed that under certain conditions the  boojums of P particles have a split core structure with broken axial symmetry \cite{Silvestre.inprep}. In this case, the P particle induces a biaxial quadrupole \cite{Pergamenshchik.2010} thus adding a correction to the asymptotic result, Eq.\eqref{QQ}. However, since the core size is of the order of the nematic correlation length, the effects of the biaxiality should be small far away from the particle. Indeed, as we shall demonstrate below, our calculations are in good agreement with Eq.\eqref{QQ} indicating that the biaxial contribution is small.

We calculate the two-body effective interaction potential $F^{(2)}(d,\theta)$
\cite{Tasinkevych.2006} which is defined by decomposing the excess 
(over the free energy of the uniform nematic) free energy for two particles $F_2\equiv F - f_b(Q_b)\Omega$ as follows
\begin{equation}
 F_2(d,\theta)= F^{(1)}_S + F^{(1)}_P + F^{(2)}(d,\theta).
\label{pair_pot_def}
\end{equation}
$F^{(1)}_{S,P}$ is the excess free energy calculated independently for an isolated S, P particle, respectively.
By definition, $F^{(2)}$ tends to zero when $d\rightarrow \infty$.

\begin{figure}[t]  %{r}{8cm} % "l" or "r" for the side on the page.
\centerline{\includegraphics[width=\columnwidth,angle = 0]{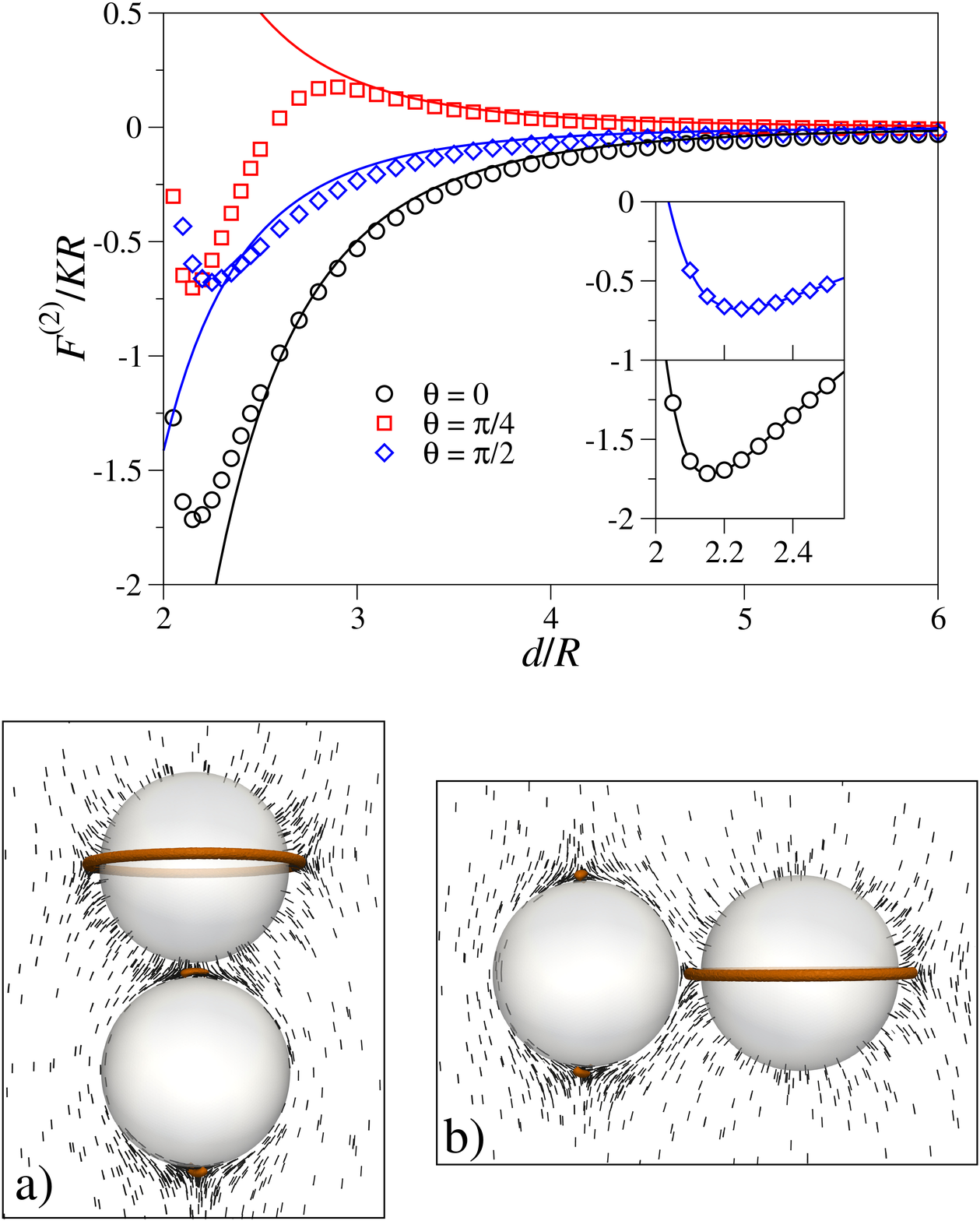}}
\caption
{Pair interaction potential $F^{(2)}$ as a function of inter-particle separation $d$ for different orientations $\theta=0,\,\pi/4,\,\pi/2$ relative to the far-field director (see Fig.\ref{dx}).  Colored lines represent the asymptotic result, Eq.\eqref{QQ}, with $Q_SQ_P=-0.18 R^6$.  The inset shows in more detail the global ($\theta=0$) and local ($\theta=\pi/2$) minima located at $d=d_\parallel\simeq2.16R$ and $d=d_\perp\simeq2.25R$, respectively. The nematic configurations are illustrated in a) for the global minimum at $\theta=0, d=d_\parallel$  and in b) for the local minimum at $\theta=\pi/2,d=d_\perp $. The iso-surfaces correspond to $Q=0.25$, $Q_b \simeq 0.44$, and the short black lines represent the director field in the plane $y = 0$.}
\label{energy}
\end{figure}

\begin{figure}[t]  %{r}{8cm} % "l" or "r" for the side on the page.
\centerline{\includegraphics[width=\columnwidth,angle = 0]{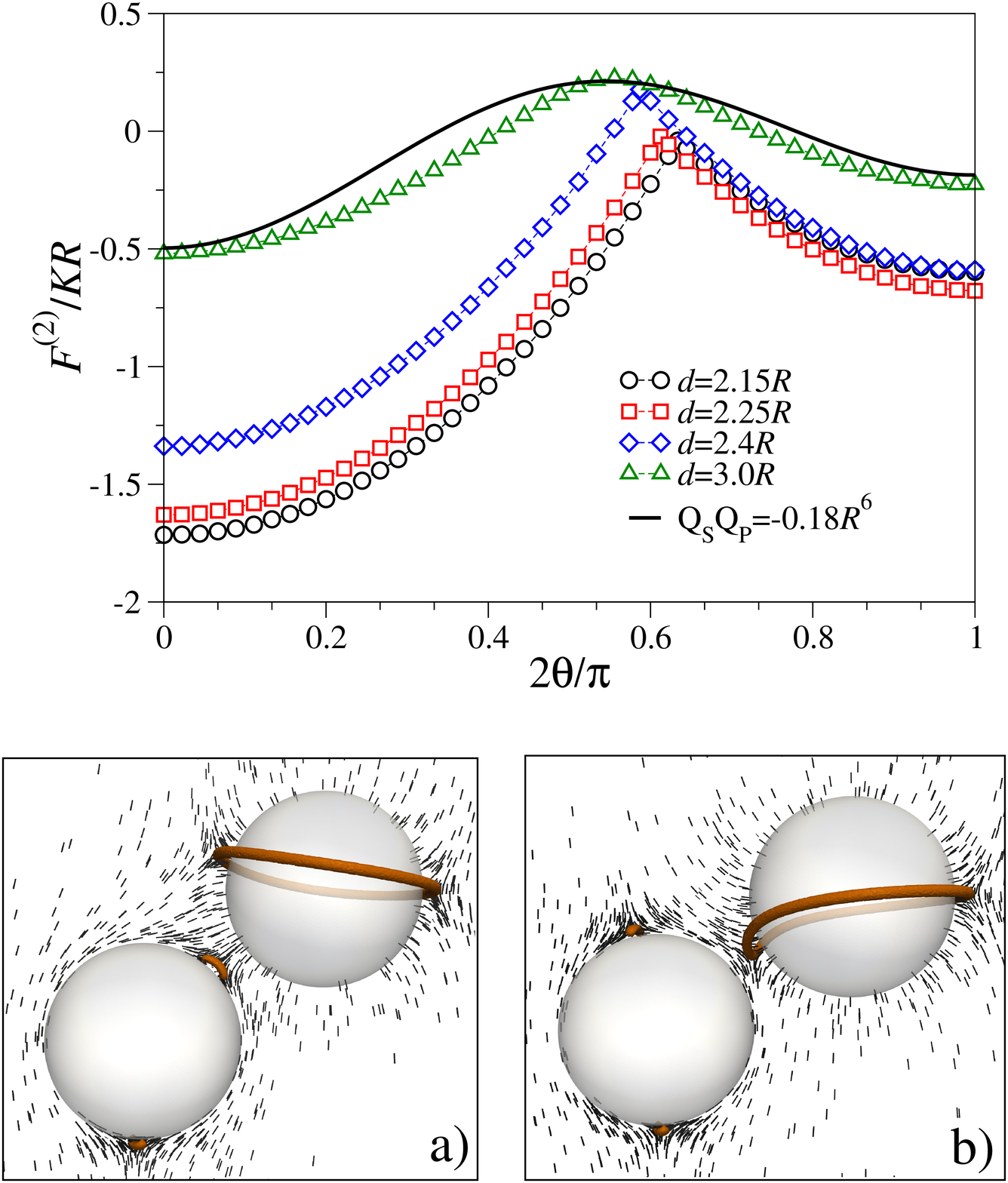}}
\caption
{Pair interaction potential $F^{(2)}$ as a function of the relative orientation $\theta$ for different inter-particle distances $d$. The black line shows the asymptotic result Eq.\eqref{QQ} with $Q_S Q_P=-0.18 R^6$ at $d = 3 R$. The snapshots depict the nematic configurations obtained from the numerical calculations at $d=2.4R$ a) $\theta\simeq0.55(\pi/2)$ and b) $\theta\simeq0.62(\pi/2)$ close to the free-energy cusp. The isosurfaces correspond to $Q=0.25$, $Q_b \simeq 0.44$, and the short black
lines represent the director field in the plane $y = 0$.}
\label{cusp}
\end{figure}

Figure \ref{energy} illustrates the S-P interaction potential $F^{(2)}(d)$ for several values of $\theta$.
By contrast to the S-S and P-P potentials, the S-P interaction is purely attractive at  $\theta=0,\,\pi/2$, and repulsive at intermediate orientations, which is consistent with $Q_SQ_P<0$. As shown in Fig.~\ref{energy}, the asymptotic result Eq.\eqref{QQ} is obeyed accurately for $d\gtrsim3R$. For smaller distances the superposition approximation is qualitatively wrong, indicating the significance of non-linear effects. $F^{(2)}(d,\theta)$ exhibits a local minimum at $d_\perp\simeq2.25R$, corresponding to $\theta = \pi/2$, and a global minimum at $d_\parallel\simeq2.16R$ for relative orientation $\theta=0$ (see also Fig.~\ref{cusp}). These values are in excellent agreement with the experimental results $d_\perp^{exp}\simeq2.28R$ and $d_\parallel^{exp}\simeq2.16R$ \cite{Ognysta.2011}. 
At $\theta=\pi/4$, $F^{(2)}(d)$ reveals a crossover from repulsive, at large $d\gtrsim 2.9 R$, to attractive behaviour, at smaller $d \lesssim 2.9 R$.  
The crossover is driven by the structural re-arrangement of the defects (see the configurations in Fig.~\ref{cusp}a) and b) below). This is similar to the crossover observed in the S-S interaction \cite{Tasinkevych.2010} at $\theta=\pi/2$, where the Saturn-rings deform from their equatorial position, or for the P-P interaction at $\theta=0$, where the inner boojums slide on the colloidal surfaces in opposite directions \cite{Silvestre.inprep,Mozaffari.2011}. 
In the S-P interaction potential, however, we observe both the sliding of the inner boojum and the deformation of the Saturn-ring, as shown in Fig.~\ref{cusp} below. 
The configuration at $\theta=\pi/4,\, d\simeq2.14 R$ does not correspond to a local minimum of the S-P interaction potential. In fact, $F^{(2)}(d,\theta)$ in the vicinity of that point has a cusp-like shape in the $\theta-$direction (see Fig.~\ref{cusp}). At very small distances, however, the conflicting anchoring conditions on the particle surfaces lead to a repulsive $F^{(2)}(d)$ at any orientation $\theta$. 

\begin{figure}[t]  %{r}{8cm} % "l" or "r" for the side on the page.
\centerline{\includegraphics[width=\columnwidth,angle = 0]{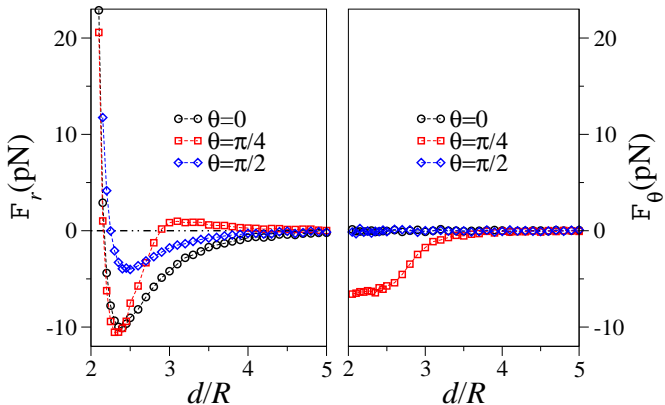}}
\caption
{Radial $\mathtt{F}_r$ and polar $\mathtt{F}_\theta$ components of the force acting on the colloidal particles as a function of the inter-particle distance $d$, for several orientations $\theta=0,\,\pi/4,\,\pi/2$ relative to the far-field director $\nvec_\infty$.}
\label{force}
\end{figure}

This can be seen also from the behaviour of the elastic force $\mathbf{F}(d,\theta)$ between the colloidal particles. Figure \ref{force} shows the radial $\mathtt{F}_r=-\partial F^{(2)}/\partial d$ and the polar $\mathtt{F}_\theta=-(1/d)\partial F^{(2)}/\partial \theta$ components of the force as functions of $d$ for several values of $\theta$. It is clear that for the orientations $\theta=0$ and $\theta=\pi/2$ the interaction is purely attractive at large separations $d$, and that in all directions a strong repulsion appears at very small $d$, stabilizing the dimer at a fixed separation $d>2 R$. Moreover, the behaviour of $\mathtt{F}_\theta$ clearly shows that once the particles are oriented along $\theta=0$ or $\theta=\pi/2$ they will approach each other in this directions under the action of $\mathtt{F}_r$, and will stabilize at $d=d_{\parallel}$ or $d = d_{\perp}$, respectively. The radial component of the force at $\theta=\pi/4$ is repulsive for large inter-particle distances, and turns into attractive as soon as the defects start to rearrange. Although, the asymptotic result Eq.\eqref{QQ} predicts a non-zero $\mathtt{F}_\theta<0$ at $\theta=\pi/4$, i.e., a polar component of the force which tends to align the particles parallel to $\nvec_\infty$, it is only when the non-linear elastic effects become important ($d\lesssim3R$) that $\mathtt{F}_\theta$ increases significantly, and becomes comparable to $\mathtt{F}_r$. This is in sharp contrast to the P-P interaction force \cite{Silvestre.inprep}, where $\mathtt{F}_\theta$ is always two orders of magnitude smaller than $\mathtt{F}_r$.

Figure \ref{cusp} illustrates the interaction potential $F^{(2)}(\theta)$ for several values of the inter-particle distance $d$. At $d=3.0R$ the interaction potential is a smooth function of $\theta$. It is clear that for this distance the asymptotic result, Eq.\eqref{QQ}, represented in Fig.~\ref{cusp} using the experimental value $Q_SQ_P=-0.18 R^6$, is qualitatively but not quantitatively accurate. As mentioned previously, the asymptotic result describes the full LdG results quantitatively at distances $d>3R$. As the distance between the particles decreases, $F^{(2)}(\theta)$ reveals a cusp at some oblique angle $\theta$. This corresponds to a structural transition between two different nematic configurations, which ``coexist'' at the orientation corresponding to the cusp. Two representative nematic configurations are shown at $d=2.4R$ in Fig.~\ref{cusp}a) at $\theta\simeq 0.55(\pi/2)$ and b) at $\theta\simeq0.62(\pi/2)$. When the S and P colloidal particles are at the equilibrium configuration at $\theta=0$, increasing $\theta$ forces the inner boojum to slide in the same direction as a result of the repulsion from the disclination ring, which deforms upwards from its equatorial position, as depicted in Fig.~\ref{cusp}a). The S-P system follows the left branch of $F^{(2)}(\theta)$ as $\theta$ increases from zero. This configuration becomes metastable at the cusp and at a larger value of $\theta^*(d)$ the repulsion between the inner boojum 
and the disclination ring renders it unstable. Then, the system jumps to the right branch, and assumes a configuration similar to that depicted in Fig.~\ref{cusp}b). On the other hand, 
when the S-P system starts from the local equilibrium configuration at $\theta=\pi/2$, it follows the right branch of $F^{(2)}$ as $\theta$ decreases. The disclination ring now deforms downwards and pushes one of the boojums away from the $z-$axis as depicted in Fig.~\ref{cusp}b). 

\begin{figure}[t]  %{r}{8cm} % "l" or "r" for the side on the page.
\centerline{\includegraphics[width=\columnwidth,angle = 0]{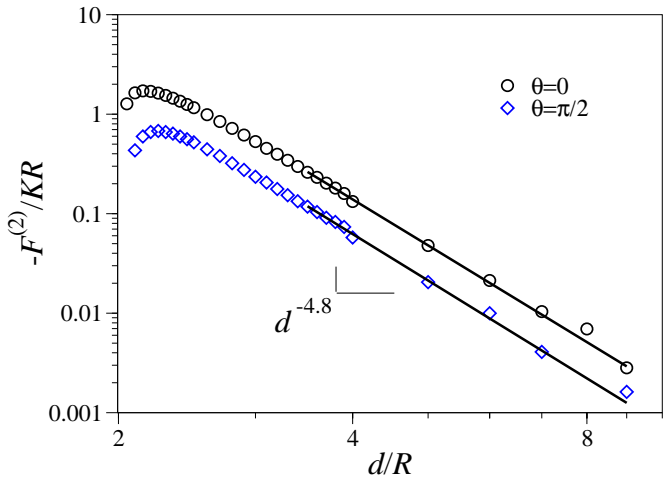}}
\caption
{Pair interaction potential $F^{(2)}$ as a function of the inter-particle distance $d$ for $\theta = 0,\,\pi/2$. Black lines correspond to a power law fit $F^{(2)}=f(\theta)\left(R/d\right)^{\alpha(\theta)}$ to the numerical results. In both cases $\alpha=4.8\pm 0.2$.}
\label{loglog}
\end{figure}

It is clear from Figs.~\ref{energy} and \ref{cusp}, that at short-distances $d\lesssim3R$ the inter-particle interaction deviates significantly from the asymptotic result described by Eq.\eqref{QQ}. In Fig.~\ref{loglog} we plot the absolute value of the S-P  interaction potential  as a function of $d$ for $\theta=0,\,\pi/2$. For large separations, $d \gtrsim 3R$,  $F^{(2)}(d)$ can be described by a power law
\begin{equation}
F^{(2)}=f(\theta)(R/d)^{\alpha(\theta)},
\label{fit}
\end{equation} 
where both $\alpha(\theta)$ and $f(\theta)$ are fitting parameters. Figure \ref{energy} indicates that the asymptotic result is valid for $d\gtrsim3R$. Thus, in Fig.~\ref{loglog} we 
fit Eq.\eqref{fit} to data corresponding to $d\geq3.5R$ only. For both values of $\theta$ we find $\alpha=4.8\pm 0.2$, which is in good agreement with the asymptotic power law $\propto1/d^5$. From these fits we also obtain  $Q_SQ_P/R^6=-0.15 \pm 0.03$ and $Q_SQ_P/R^6=-0.20\pm0.05$ for $\theta=0$ and $\theta=\pi/2$, respectively. This is in excellent 
agreement with the experimental value $Q_SQ_P/R^6=-0.18$ reported in Ref.~\cite{Ognysta.2011}.

\section{Conclusions}
\label{conclusions}
Recently, a directed assembly of spherical particles dispersed in a nematic host  into $2D$ square crystal-like structures was reported  \cite{Ognysta.2011}. This has been achieved by mixing quadrupolar colloidal particles with distinct anchoring conditions, namely homeotropic and degenerate planar. The homeotropic anchoring particles nucleate Saturn-ring defects, and the planar anchoring particles antipodal boojums. In the linear regime each of these colloidal particles induce long-range distortions of the nematic director with quadrupolar symmetry \cite{Lev.2002}, and as a consequence the long-range pair interaction is of the quadrupolar type. In this paper we have studied numerically the nematic-mediated interaction between two such particles. For simplicity we have assumed spherical particles of equal radii.  

We have confirmed that the S and P particles have quadrupole moments with opposite signs, $Q_SQ_P<0$. The asymptotic analysis \cite{Pergamenshchik.2010} suggests that S particles have a  positive quadrupole moment, and  P particles a negative one. At large distances the interaction potential $F^{(2)}(d,\theta)$ behaves as  $d^{-\alpha}$ with $\alpha=4.8\pm 0.2$ which is in excellent agreement with the asymptotic result $d^{-5}$ for quadrupolar interactions.  We have also confirmed numerically the experimental values for the product of the quadrupole moments $Q_SQ_P=-0.18 R^6$. Note that if the confining surfaces of the experimental setup \cite{Ognysta.2011} had an influence on the pair interaction, there would be a deviation from the power law behaviour that would eventually (for very strong confinement) turn into an exponential decay. Since in our calculations the finite size effects are negligible, our results indicate that in the experiment of  Ognysta et al. \cite{Ognysta.2011} the confining surfaces played a negligible role. For distances $d\lesssim 3R$, the asymptotic quadrupolar result Eq.\eqref{QQ} is qualitatively wrong, see Fig.~\ref{energy}. At such distances the topological defects start to interact and re-arrange their positions (see Fig.~\ref{cusp} a), b)),
resulting in profound changes of the interaction potential $F^{(2)}(d)$ (see Fig.~\ref{energy} red squares).

Under some conditions, the P particles nucleate boojums of complex biaxial structure \cite{Silvestre.inprep}. This leads to a biaxial quadrupolar distortion, since it breaks the axial symmetry, and could change the asymptotic uniaxial expression given by Eq.~\eqref{QQ} \cite{Pergamenshchik.2010}. However, since the size of the biaxial regions is of the order of the nematic correlation length, the biaxial perturbation at large distances  are  negligible, which  explains the good agreement between our numerical results and the asymptotic expression for uniaxial quadrupoles. 

By contrast to the interaction of particles with identical anchoring conditions (P-P, or S-S), the S-P interaction potential $F^{(2)}(d,\theta)$ exhibits two minima (in the first quadrant), when the particles are oriented at $\theta=0$ or $\theta=\pi/2$ relative to the far-field director, and separated by $d_\parallel\simeq 2.16R$ and $d_\perp\simeq 2.25R$, see Fig.~\ref{energy}. These values are in excellent agreement with those reported experimentally \cite{Ognysta.2011}, $d_\parallel^{exp}\simeq2.16R$ and $d_\perp^{exp}\simeq 2.28R$  \cite{Ognysta.2011}. We have also found a strong repulsion at short distances and at any orientation $\theta$. These two features of the pair interaction potential are pre-requisites for the assembly of colloidal lattices with rectangular unit cells. 

We have calculated the pair elastic force, see Fig.~\ref{force}. The polar component $\mathtt{F}_\theta$ is always zero at $\theta = 0,\,\pi/2$, and acquires rather large values at oblique orientations $\theta$,  provided that the inter-particle distance is not too large ($d \lesssim 3 R$, see Fig.~\ref{force} right panel). The sign of $\mathtt{F}_\theta$
depends on $\theta$. $\mathtt{F}_\theta < 0$ for $\theta \leqslant \theta^c(d)$, and it is positive otherwise.
$\theta^c(d)$ is the angular position of the cusp in the $F^{(2)}(\theta)$ profile, see Fig.~\ref{cusp}. Therefore, there is a
strong elastic force which aligns a pair of colloidal particles either parallel ($\theta \leqslant \theta^c(d)$), or perpendicular  
($\theta \geqslant \theta^c(d)$) to the far-field director. 
Once the particles are aligned along one of these directions, they move under the action of the radial force $\mathtt{F}_r$ towards the global $d = d_\parallel, \theta =0$, or the local $d =d_\perp, \theta = \pi/2$ equilibrium positions. The presence of a strong aligning force $\mathtt{F}_\theta$ at small-to-intermediate distances is in sharp contrast with the P-P interaction profile \cite{Silvestre.inprep}, where the polar component $\mathtt{F}_\theta$ is always two orders of magnitudes smaller than the radial force  $\mathtt{F}_r$.

Most theoretical studies of the interaction between colloidal particles in nematic liquid crystals focus on particles with the same size. Although this may seem a rather strong simplification, it is justified by the current experimental ability of producing monodispersed colloidal particles. This seems to be the best choice for the assembly of photonic crystals, but the effect of polydispersity on the stability of such crystals remains an important open question.

\section*{Acknowledgments}

We gratefully acknowledge financial support from the Portuguese Foundation for Science and Technology (FCT) through Grants Nos. PEstOE/FIS/U10618/2011, PTDC/FIS/098254/2008, SFRH/BPD/40327/2007 (NMS) We also acknowledge partial financial support from FCT-DAAD Transnational Cooperation Scheme under the Grant  No. 50108964, and the FP7 IRSES Marie-Curie grant PIRSES-GA-2010-269181.

\footnotesize{
\bibliography{all}
}
\end{document}